\documentclass[12pt]{article}
\usepackage{amsmath, latexsym, amssymb}

\numberwithin{equation}{section}
\newcommand{\R}{\mathbb R}
\newcommand{\C}{\mathbb C}

\newcommand{\X}{\mathbb{X}}

\newcommand{\N}{\mathbb N}

\newcommand{\K}{K}
\newcommand{\p}{\partial}

\newcommand{\Ta}{T_0}

\newcommand{\Bold}[1]{{\boldsymbol{\mathit{#1}}}}

\newcommand{\ti}{\mathrm{t}}

\newcommand{\x}{\mathrm{x}}
\newcommand{\s}{\mathrm{s}}
\newcommand{\g}{\mathrm{g}}
\newcommand{\xx}{\bar{x}}

\newcommand{\m}{\mathrm{m}}

\newcommand{\M}{\mathbb{M}}
\newcommand{\U}{\mathbb{U}}
\newcommand{\I}{\mathbb{I}}

\newcommand{\PP}{\mathbb{P}}

\newcommand{\QED}{\hspace{.2in}\square\newline}

\newtheorem{claim}{Claim}[section]

\newtheorem{proposition}{Proposition}[section]
\newtheorem{definition}{Definition}[section]

\begin{document}

\begin{center}
{\Large \textbf{Functional Integration on\\Constrained Function
Spaces II: Applications}} \vskip 2em
{ J. LaChapelle}\\
\vskip 2em
\end{center}

\begin{abstract}
Some well-known examples of constrained quantum systems commonly quantized via Feynman path integrals are re-examined using the notion of conditional integrators introduced in the companion paper \cite{LA4}. The examples yield some new perspectives on old results. As an interesting new application, the formalism is used to construct a physical model of average prime counting functions modeled as a constrained gamma process.
\end{abstract}

\vskip 1em


\vskip 1em



\section{Introduction}
A basis for functional integration on constrained function spaces was proposed in \cite{LA4} in analogy with Bayesian inference theory. Needless to say, the proposed formalism must reproduce known results. So here some implications and
applications of various integral representations are checked against four archetypical classes of constrained systems that were heuristically reviewed in \cite{LA4}. The four classes can be roughly characterized as: Kinematical; a) fixed end-points, b) bounded configuration space, c) segmented configuration space, and Dynamical; a) functional constraints.  The subsequent efficient derivation of old results illustrates the utility of the new techniques introduced.

This exercise requires the development of some new tools. In particular, kinematical constraints suggest the notion of a Dirac delta functional on the topological dual of the constraint space. It turns out that these delta functionals are particular types of gamma functional integrals. In fact, gamma functional integrals can be used to define `step functionals' and all their (Gateaux) derivatives. Presumably, one could use such tools to construct a functional analog of distribution theory. Similarly, dynamical constraints suggest the notion of Dirac integrators characterized by Dirac delta functionals on the constraint space (as opposed to the dual constraint space). Their heuristic equivalents have long been used to enforce functional constraints --- the most notable example being the Faddeev-Popov method in quantum field theory. Here we use them to treat topologically nontrivial systems that can be modeled as quotient spaces: Their associated functional constraints can be efficiently encoded using paths in an appropriate fiber bundle that are constrained to be horizontal relative to a given connection. Finally, discontinuities in configuration space can be handled using the idea of ``path decomposition" \cite{AU/SC}--\cite{HA}. The resulting functional integral tools lead to a recursive process to calculate  propagators on  bounded and/or segmented configuration spaces. The recursive process can be evaluated iteratively, but  ``path decomposition'' suggests a new and potentially useful approximation technique rooted in boundary Green's functions. More importantly, the functional integral construction shows that the propagators associated with such systems are intimately related to Poisson integrators and, hence, gamma integrators.

As a new application of the formalism, we propose functional integral
representations of some prime counting functions.  The functional integrals represent \emph{average} counting functions, and they give excellent approximations to the exact counting functions. According to the construction, the prime counting functions are modeled
as a constrained gamma process (as opposed to constrained Gaussian processes of the quantum examples)  --- a perspective that might lead to a
deeper understanding of the distribution of prime numbers. This is a nice realization of the often observed interplay between physics and number theory.

We should alert the reader that results from \cite{LA4}, particularly notation and details of integrator families presented in appendix B \cite{LA4}, will be used without explanation.

\section{Two conditional integrators}
The value of the formalism proposed in \cite{LA4} is perhaps best appreciated by
application to the familiar examples outlined in \S 2 \cite{LA4} in the context of quantum mechanics (QM). Recall that the task is to represent constrained function spaces as enlarged unconstrained function spaces--- including non-dynamical degrees of freedom --- equipped with conditional and conjugate integrators.

Now,  in simple QM, eigenfunctions of position observables for the
semiclassical approximation are characterized by their mean and
covariance. When viewed as functions of time, they become paths in
some configuration space. Consequently, semiclassical QM can be
described in terms of functional integrals with \emph{Gaussian integrators}
on the space of paths characterized by an assumed mean and
covariance.

Applying \S3 \cite{LA4} to the case\footnote{Recall that $C$ is a Banach space of non-dynamical degrees of freedom induced by constraints on a dynamical system whose degrees of freedom are elements of a Banach space $X$.} $Y\equiv C$, it is clear that the imposition of constraints in QM can change the
quadratic form and/or the mean associated with an unconstrained
Gaussian integrator. The notion of conjugate integrators enables one
to construct the `marginal' and `conditional' integrators associated
with those constraints, and hence make sense of constrained
functional integrals. We will see that the type of conjugate integrator family
depends on the type of constraint.

\subsection{Delta functional on $C'$}
\begin{claim}\label{uniform functional}
Let $X_a$ denote the space of $L^{2,1}$ maps
$x:[\ti_a,\ti_b]\rightarrow\mathbb{X}$ with $x(\ti_a)=\x_a\in\X$, and let it be endowed with a Gaussian integrator $\mathcal{D}\omega_{\xx,\mathrm{Q}}(x)$. Given the Banach space $C$ associated with a constraint on $X_a$, a delta functional on its topological dual $C'$ is induced by a constraint on the mean but not the covariance
of $x\in X_a$ relative to $\mathcal{D}\omega_{\xx,\mathrm{Q}}(x)$.
\end{claim}

According to the Bayesian analogy, we consult a table of conjugate prior probability distributions and find the conjugate family associated with a Gaussian distribution of known
mean and unknown covariance is a gamma distribution, So the
integrator on $C$ associated with fixed-mean functional constraints is expected to be a
gamma integrator.\footnote{The gamma family is also conjugate for
exponential-type distributions so the following analysis holds for
action functionals that are not necessarily quadratic.}

Now that we know the `probability' nature of fixed-mean functional constraints on a Gaussian QM system through its marginal integrator family, the task is to understand the integrator family parameters and the relevant sufficient statistics.

As motivation, consider the lower gamma integrator family (see Appendix B, \cite{LA4}). Put
$\alpha=1$, and let $L:\Ta\rightarrow i\R^n$ with $\langle
\beta',L(\tau)\rangle\mapsto 2\pi \Bold{\lambda}\cdot i
\Bold{\mathrm{u}}$ and $\Bold{\lambda}=\Bold{\lambda}^\ast$. Then,
\begin{equation}
\int_{\Ta}\mathcal{D}\gamma_{1,\beta',\infty}(\tau)\stackrel{L}{\longrightarrow}
\int_{\R^n}e^{-2\pi i\Bold{\lambda}\cdot
\Bold{\mathrm{u}}}d\Bold{\mathrm{u}}=\delta(\Bold{\lambda})\;.
\end{equation}
with $\Bold{\lambda}\in\R^n$ and the integral over $\R^n$ understood
as an inverse Fourier transform. On the other hand,
\begin{equation}
\int_{{\Ta}} \mathcal{D}\gamma_{1, \beta',\tau_o}(\tau)
:=
\frac{\gamma(1,\tau_o)}{ \mathrm{Det}(\beta')}
=\frac{1-e^{- \tau_o}}{ \mathrm{Det}(\beta')}\;,
\end{equation}
and so the integrator $\mathcal{D}{\gamma}_{1,\beta',\infty}(\tau)$ can be understood as a limit;
\begin{equation}
\int_{\Ta}\mathcal{D}{\gamma}_{1,\beta',\infty}( \tau): =\lim_{
|\tau_o|\rightarrow\infty}
\int_{\Ta}\mathcal{D}{\gamma}_{1,\beta',\tau_o}( \tau)\;.
\end{equation}

Consequently, when $\tau_o$ is strictly imaginary,
$\mathcal{D}\gamma_{1,i\beta',\infty}(\tau)$ can be interpreted as
the functional analog of a two-sided Laplace transform implying
\begin{equation}
\int_{\Ta} \mathcal{D}\gamma_{1, i\beta',\infty}(\tau) =\lim_{
|\tau_o|\rightarrow\infty} \frac{e^{
\tau_o}-e^{- \tau_o}} {\mathrm{Det}(i\beta')}\;;
\end{equation}
which formally vanishes except when $\mathrm{Det}(i\beta')=0$. In fact, this can be interpreted as the functional analog of a delta function. In particular, this integrator can be used to localize onto a subset of the dual constraint space $C'$.

This justifies the definition:
\begin{definition}\label{def 4.4}
Suppose $i\langle c',c\rangle\in\I:=i\R$ and $S_s(c')$ is degenerate on $C$. A delta functional\footnote{Since $C'$
is a polish space, the delta functional can be interpreted in a
measure theoretic sense as the complex Borel measure associated with
the identity element in the Banach space $\mathbf{F}(C)$, i.e.
$\mathrm{Id}=\mathrm{F}_{\mu_\delta}(c)=\int_{C'}e^{ i\langle
c',c\rangle}d\mu_\delta(c')\;\;\forall c\in C$.} on $C'$ is defined by
\begin{equation}
\delta_{S_s}(c')
:=\frac{1}{\Gamma(1)}\int_{C}\mathcal{D}{\gamma}_{1,ic',\infty}(c)
\end{equation}
and a Heaviside step functional by
\begin{equation}
\theta_{S_s}(c'):=\frac{-i}{\Gamma(0)}\int_{C}
\mathcal{D}{\gamma}_{0,ic',\infty}(c)\;.
\end{equation}
\end{definition}
The subscript $S_s$ reminds that the functionals are evaluated on the sufficient statistic subspace. Consequently, the delta functional vanishes unless $\mathrm{Det}S_s(c')=0$, and the step
functional vanishes unless $\mathrm{Det}S_s(c')\in\C_+$.

Remark that this definition suggests the
characterization
\begin{equation}
\delta_{S_s}^{(\alpha-1)'}(c')
=\frac{i^{\alpha-1}}{\Gamma(\alpha)}\int_{C}\mathcal{D}{\gamma}_{\alpha,ic',\infty}(c)
\end{equation}
when $i\langle c',c\rangle\in\I$ and $\mathrm{Det}S_s(c')=0$.
 The characterization is ``good" in the sense that $\delta_{S_s}(c')$ reduces to the usual Dirac delta function under linear maps ${L}:C\rightarrow\mathbb{R}^n$ for any $n$, and for $\alpha=m\geq1$ with $m\in\N$ we have
 \begin{equation}
 \delta_{S_s}^{(m-1)'}(c')(\ti)=i^{m-1}\frac{\Gamma(m-1)}{\Gamma(m)}
 \int_{C}\frac{\delta^m}{\delta c'(\ti)^m}\mathcal{D}{\gamma}_{0,ic',\infty}(c)\;.
 \end{equation}
Evidently gamma integrators and their associated functional
integrals could be used as a basis for a theory of what might be
called `distributionals'.

\subsection{Delta functional on $C$}

\begin{claim}\label{def4.2}
Let $X_a$ be endowed with a Gaussian
integrator $\mathcal{D}\omega_{\xx,\mathrm{Q}}(x)$. A delta functional on
$C$ represents a constraint on the covariance but not the mean
of $x\in X_a$ relative to $\mathcal{D}\omega_{\xx,\mathrm{Q}}(x)$.
\end{claim}

Again consulting a table of conjugate priors, the conjugate family associated with a Gaussian distribution of known
covariance and unknown mean is again a Gaussian distribution so the
integrator associated with a fixed-covariance functional constraint is
expected to be a Gaussian. Essentially this means that the unknown
mean of the conditional integrator on $X_a$ is normally distributed
with respect to the marginal \emph{and} conditional integrators on
$C$.

\begin{definition}
A Dirac integrator\footnote{The Dirac integrator is improper in the
sense that it is a limit of a Gaussian that requires regularization
to achieve a sensible normalization.} is defined by
\begin{equation}
\mathcal{D}\delta_{\bar{c}}(c):=\mathcal{D}\omega_{\bar{c},\infty}(c)
=:\delta(c-\bar{c})\mathcal{D}c\;\;\;\mathrm{such\;
that}\;\;\;\lim_{\mathrm{W}(c')\rightarrow 0}\mathrm{Det}(\mathrm{W})^{1/2}e^{-\pi
\mathrm{W}(c')}:=1\;.
\end{equation}
\end{definition}

\begin{proposition}
The Dirac integrator is normalized
$\int_{C}\mathcal{D}\delta_{\bar{c}}(c)=1$, translation invariant
$\mathcal{D}\delta_{\bar{c}}(c-c_0)=\mathcal{D}\delta_{\bar{c}}(c)$,
and furnishes the functional analog of a Dirac measure
\begin{equation}
\int_{C}\mathrm{F}_{\mu}(c)\mathcal{D}\delta_{\bar{c}}(c)=\mathrm{F}_{\mu}(\bar{c})\;.
\end{equation}
\end{proposition}

\emph{Proof.} The normalization is obvious and the translation
invariance follows from the translation invariance of the primitive
integrator. The third relation follows trivially from definitions and translation invariance of the primitive Gaussian integrator;
\begin{eqnarray}
\int_{C}\mathrm{F}_{\mu}(c)\mathcal{D}\delta_{0}(c)
&=&\int_{C'}\mathrm{Z}(c')\;d\mu(c')\notag\\ &=&\int_{C'}1\;d\mu(c')\notag\\
&=&\mathrm{F}_{\mu}(0)\;.
\end{eqnarray}
The last line follows from the definition of $\mathrm{F}_{\mu}(c)$ and $c=\bar{c}\Rightarrow e^{\pi
i\langle c',c-\bar{c}\rangle}=1\;\forall c'\in C'$.
$\QED$

Justification for the term `delta functional' is obvious: Similar to the gamma integrator with $\alpha=1$ characterizing the identity on $C$, the Dirac integrator characterizes the identity on $C'$ in the sense that
$\int_{C}e^{-2\pi i\langle c',c\rangle}\mathcal{D}\delta_{\bar{c}}(c)=\mathrm{Id}$ for all $c'\in C'$.

To see that Dirac integrators are conjugate integrators, use eq.
(B.18) \cite{LA4} in the context of sufficient
statistics. To make this concrete we stipulate that
\begin{equation}\label{sufficient statistics}
G_{cx}^{-1}=D_{cx}:S_s(X_a)\rightarrow
C'\;\;\;;\;\;\;G_{xc}^{-1}=D_{xc}:C\rightarrow S_s(X_a)'\;.
\end{equation}
Insofar as the sufficient statistics characterize `classical'
observables in a QM context, this means the constraints are only
correlated with classical observables. Now taking the
$G_{cc}\rightarrow\infty$ limit gives
\begin{equation}
e^{-(\pi/\s)\left[ \mathrm{Q}_C(c-m_{c|x})-\mathrm{B}(m_{c|x})\right]}\longrightarrow
e^{-(\pi/\s)\left[\widehat{\mathrm{Q}}(c-m_{c|x})-\mathrm{B}(m_{c|x})\right]}\delta(c-m_{c|x})
\end{equation}
where
$\widehat{\mathrm{Q}}(c_1,c_2)=\langle(G_{cx}D_{xx}G_{xc})^{-1}c_1,c_2\rangle=:
\langle\widehat{G}^{-1}c_1,c_2\rangle$ and
$m_{c|x}=\bar{c}+G_{cx}D_{xx}(x-\xx)$.

Clearly,
$\widehat{\mathrm{Q}}(c_1,c_2)=\left.\mathrm{Q}_{X}(x_1,x_2)\right|_{S_s(X_a)}$
because of (\ref{sufficient statistics}) so the exponential can be
viewed as a likelihood functional
\begin{equation}
\Theta_{X|C}(S_s(x)|c,\cdot)=e^{-(\pi/\s)\left[ \mathrm{Q}_X(c-\bar{x})-\mathrm{B}(\xx)\right]}
\end{equation}
where now $\bar{x}:=(D_{xx}G_{xc})^{-1}m_{c|x}$. It follows that
\begin{equation}
\Theta_{C|X}(c|x,\cdot)\sim e^{-(\pi/\s)\left[\mathrm{Q}_X(c-\bar{x})-\mathrm{B}(\xx)\right]}\delta_{\bar{c}}(c)
\end{equation}
and, hence,
\begin{equation}
\mathcal{D}\omega_{m_{c|x},\mathrm{Q}_{C|X}}(c|x)\propto e^{-(\pi/\s)\left[\mathrm{Q}_X(c-\bar{x})-\mathrm{B}(\xx)\right]}\delta_{\bar{c}}(c)\mathcal{D}(x,c)\;.
\end{equation}
Since the mean is altered but not the covariance, the pair
$(\mathcal{D}\omega_{m_{c|x},\mathrm{Q}_{C|X}}(c),\mathcal{D}\omega_{\bar{c},\infty}(c))$
are conjugate integrators and
$\int_{C}\delta(c-\bar{c})\mathcal{D}c$ enforces a delta functional
constraint.

More generally, the Dirac integrator yields
\begin{equation}
\int_{C}\mathrm{F}_{\mu}(c)\mathcal{D}\delta_{\bar{c}}(M(c))
=\sum_{c_0}\frac{\mathrm{F}_{\mu}(c_0)}{\mathrm{Det}M'_{(c_0)}}
\end{equation}
with $M:C\rightarrow C$ a diffeomorphism, $M(c_0-\bar{c})=0$, and $\mathrm{Det}M'_{(c_0)}$ non-vanishing and appropriately regularized. Nothing is altered if we allow general action functionals. That is,
we can define Dirac integrators as the limit $\mathrm{S}(c)\rightarrow 0$ for
exponential-type integrators. The quintessential example of this
type (which, however, is outside the scope of this article) is gauge
fixing in quantum field theory. Assuming well-defined functional
integrals for fields, a Dirac integrator on the space of gauge transformations $G$ is just the Faddeev-Popov
`trick'
\begin{equation}
\int_G\mathcal{D}\delta_{\bar{g}}(M(g))=\sum_{g_o}
\frac{1}{\mathrm{Det}M'_{(g_o)}}
\end{equation}
where $M:G\rightarrow G$ and $M(g_o-\bar{g})(a)=0$ with $a\in A$. Here $A$ is the space of connections on some principal bundle. Of course, an
admissible gauge fixing condition requires a single $g_o$, and it is
standard to average the delta functional over $\bar{g}$ with respect
to some (usually Gaussian) integrator.

The whole Faddeev-Popov procedure can be readily interpreted from a
conditional integrator viewpoint (eq. (3.14) \cite{LA4}). Formally,
\begin{eqnarray}
\int_{\widetilde{A}} \mathrm{F}(\widetilde{a})e^{i\mathrm{S}(\widetilde{a})}\mathcal{D}(\widetilde{a})
&=&\int_{A\times G}\left[\mathrm{F}(a)\,\mathrm{Det}\left(M'_{g_o}(a)\right)\,\delta(M(g_o(a)))\;\mathcal{D}(g)\right]\,e^{i\mathrm{S}(a)}\;\mathcal{D}(a)
\notag\\
&=:&\int_A \widetilde{\mathrm{G}}(a)\,e^{i\mathrm{S}(a)}\mathcal{D}(a)\;.
\end{eqnarray}
where $\widetilde{\mathrm{G}}(a)$ absorbs the gauge group volume.

\section{Examples revisited}\label{examples}

\subsection{Fixed end-points}
Return to the free QM point-to-point propagator $K(x_a,x_b)$ in $\R^n$ discussed in  \S 2 \cite{LA4}. The constraint is clearly gamma-type. Define
\begin{equation}
 i\langle c'(x),c\rangle:=2\pi i\int_{\ti_a}^{\ti_b}\left|x(\ti)-\bar{x}(\ti)\right| c(\ti)\;d\ti
\end{equation}
where $c(\ti)\in\R$. Impose the constraint $\delta(x(\ti_b)-\x_b)$ by choosing sufficient statistics according to
\begin{eqnarray}
-i\langle S_s(c'(x)),c\rangle&:=& 2\pi i
\int_{\ti_a}^{\ti_b}\left|x(\ti_b)-\bar{x}(\ti_b)\right| c(\ti)\;dt\notag\\
&=&2\pi i
\left|x(\ti_b)-\bar{x}(\ti_b)\right|\int_{\ti_a}^{\ti_b}c(\ti)\;dt\notag\\
&=:&2\pi i \left|x(\ti_b)-\x_b\right|\bar{c}\notag\\
&\equiv&0\;\;\forall c\in C\;.
\end{eqnarray}
Being a Banach space, the vanishing of $\left|x(\ti_b)-\bar{x}(\ti_b)\right|$ ensures  $x(\ti_b)=\bar{x}(\ti_b)$.

We have $\bar{x}(\ti_b)=\x_a+\dot{\x}_b(\ti_b-\ti_a)=\x_b$ in the third line because the mean
(actually critical in this case) path in $X_a$ is given by
\begin{equation}
\xx(\ti)=\frac{\x_a(\ti_b-\ti)}{(\ti_b-\ti_a)}+{\dot{\x}}_b(\ti-\ti_a)\;.
\end{equation}
On the other hand, the space of point-to-point paths $X_{a,b}$ has boundary conditions
$\xx(\ti_a)={\x}_a$ and $\xx(\ti_b)=\x_b$. That is, the mean path is alternatively parametrized as
\begin{equation}
\xx(\ti)=\frac{\x_a(\ti_b-\ti)+{\x}_b(\ti-\ti_a)}{(\ti_b-\ti_a)}\;.
\end{equation}

Clearly, the
integral in Definition \ref{def 4.4} reduces to a simple delta
function $\delta(|x(\ti_b)-\x_b|)$ on $\R$  for this choice of sufficient statistics since the integrand is a \emph{function} of $\bar{c}$, and we can write
\begin{eqnarray}
\int_{X_{a,b}}\mathcal{D} \omega_{\xx,\mathrm{Q}^{(a,b)}}(x)
&:=&\int_{X_a\times
C}e^{-i\langle S_s(c'(x)),c\rangle}
\mathcal{D}_{1,0,\infty}(c)\mathcal{D} \omega_{\xx,\mathrm{Q}}(x)\notag\\
&=&\int_{X_a\times
\R}e^{2\pi
i\left|x(\ti_b)-\x_b\right|\,\bar{c}}
\;d\bar{c}\;\mathcal{D} \omega_{\xx,\mathrm{Q}}(x)\notag\\
&=&\int_{X_a}\delta(x(\ti_b),\x_b)\;\mathcal{D} \omega_{\xx,\mathrm{Q}}(x)
\end{eqnarray}
where the right-hand side is to be interpreted as the integral of a
conditional integrator on $X_a\times {C}$.

Now, instead of solving the constraint first,\footnote{Although the constraint-first calculation has been done many times
by time slicing, semi-classical, or linear mapping techniques; the
point of changing integration order here is to demonstrate that the as-defined
functional integral tools allow the calculation to be carried out
entirely at the level of the function space --- the target manifold
only makes an appearance through the boundary conditions imposed on
the mean and covariance and the regularization/normalization of the
functional determinant.} we will do the integral over $X_a$ first;
\begin{eqnarray}
K(x_a,x_b)=\int_{X_{a,b}}\mathcal{D} \omega_{\xx,\mathrm{Q}^{(a,b)}}(x)&=&\int_{X_a\times
{C}} c\,e^{- i\langle S_s(c'(x)),c\rangle}
\mathcal{D}c\mathcal{D} \omega_{\xx,\mathrm{Q}}(x)\notag\\
&=&\int_{C} \int_{X_a}  c\, e^{ 2\pi i\big\langle\big\langle
\delta_{\ti_b},|(x-\xx)|\big\rangle',\;c\big\rangle}
\mathcal{D} \omega_{\xx,\mathrm{Q}}(x)\mathcal{D}c \notag\\
&=&\int_{C} c\left[\int_{X_0} e^{ 2\pi
i\big\langle |\langle\delta_{\ti_b},\tilde{x}\rangle'|,\;c\big\rangle}e^{\pi i \mathrm{B}(\xx)}\;
\mathcal{D} \omega_{0,\mathrm{Q}}(\tilde{x})\right]\mathcal{D}c \notag\\
&=&e^{\pi i \mathrm{B}(\xx)}{\sqrt{\mathrm{det}\left[i\Bold{G}(\ti_b,\ti_b)\right]}}
\int_{C} c\, e^{\pi i\left\langle
|\mathrm{W}(\delta_{\ti_b})'|,\,c\right\rangle}
\mathcal{D}c \notag\\
&=& e^{\pi i
\mathrm{B}(\xx)}{\sqrt{\mathrm{det}\left[i\Bold{G}(\ti_b,\ti_b)\right]}}
\frac{1}{\mathrm{det}\left[|i\Bold{G}(\ti_b,\ti_b)|\right]}
\end{eqnarray}
where the second line uses functional Fubini, the third line follows using results from Appendix B \cite{LA4} to shift the integration variable, and the fourth line follows because $\mathrm{Det}(i
\mathrm{W}(\delta_{\ti_b}))=\mathrm{det}\left[i\Bold{G}(\ti_b,\ti_b)\right]\neq 0$. It should be emphasized that $\Bold{G}(\ti_b,\ti_b)$ is the
covariance matrix associated to paths $x\in X_a$ with
$x(\ti_a)=0$. The boundary form, of course, evaluates to $\mathrm{B}(\bar{x})=\dot{\x}^2_b(\ti_b-\ti_a)$ which becomes $\mathrm{B}(\bar{x})=(\x_b-\x_a)^2/(\ti_b-\ti_a)$ when expressed in terms of $\bar{x}\in X_{a,b}$.\footnote{Significantly, we do not
have to expand about the critical path to do the calculation.
Including the mean and boundary form in the definition of the integrator automatically
handles this for us. But it does more. It tells us that, when
$\mathrm{Q}\rightarrow \mathrm{S}$ (a non-quadratic action), the `sufficient statistic' of import is not the
critical path but the mean path. So, for example, the semi-classical
expansion in terms of the mean path automatically accounts for
self-interactions. In other words, once the mean is known, the
Feynman diagram procedure (now without loop diagrams) is a way to
estimate the moments of the integrator $\mathcal{D}\omega_{\xx,\mathrm{S}}$.
This is the essence of the effective action
approach in quantum field theory. Of course, the catch is $\bar{x}$ is hard to find for generic $\mathrm{S}$.}

In the presence of boundary conditions or for less trivial geometry, there will be more than one
critical path  with the appropriate boundary
conditions. In that case, it follows from eq. (B.7) \cite{LA4} that
\begin{equation}\label{full propagator}
K_\mathrm{Q}(\x_a,\x_b)=\sum_{\xx}\frac{e^{\pi i \,\mathrm{B}(\xx)}}
{\sqrt{\mathrm{det}\left[i\Bold{G}(\ti_b,\ti_b)\right]}}
=\frac{1}
{\sqrt{\mathrm{det}\left[i\Bold{G}(\ti_b,\ti_b)\right]}}\sum_{\xx}e^{\pi i \,\mathrm{B}(\xx)}\;.
\end{equation}
The subscript $\mathrm{Q}$ has been included here to emphasize that the
propagator is determined by a covariance associated with specific
boundary conditions, and it is a sum over $\xx$  of \emph{all}
Gaussian integrators with $\xx$ having the appropriate boundary
conditions. \emph{This will be a recurring theme}: propagators/kernels are represented
by a sum over relevant parameters of an integrator family.

The free QM point-to-point propagator is a specialization of the more
general integral
\begin{equation}
K_\mathrm{S}(\x_a,\x_b)=\int_{X_a}\delta_{S_s}(c'(x))\mathcal{D}
\omega_{\xx,\mathrm{S}}(x) :=\sum_{\xx}\int_{X_{\xx_a}}\delta_{S_s}(c'(x))\mathcal{D}
\omega_{\xx,\mathrm{S}}(x)
\end{equation}
where $\mathcal{D}
\omega$ is characterized by an action functional $\mathrm{S}$ that is generically not quadratic, and the mean paths satisfy
associated boundary conditions enforced by sufficient statistics on $C'$. It is to be understood as an
integral over a conditional integrator on $B_a:=X_a\times C$.

 An explicit example of this type of integral comes from the
fixed-energy propagator on phase space often heuristically represented
by \cite{GA}
\begin{equation}
G_{\mathrm{ps}}(q_b,q_a;\mathrm{E})=\int_{(Q,P)_{a,b}}\delta(\mathrm{H}(q,p)-\mathrm{E})\mathcal{D}\omega^{(a,b)}(q,p)
\end{equation}
where $\mathcal{D}\omega^{(a,b)}(q,p)$ is an appropriately defined
point-to-point integrator on some phase space and $\mathrm{H}(q,p):=\int_{\ti_a}^{\ti_b} h(q(\ti),p(\ti))\,d\ti$.

According to the functional Fubini, the order of integration can be
interchanged for product integrators (the analog of independent
joint distributions). Then, as we saw in the example, since the gamma family is conjugate for
Gaussian likelihood functionals, the integral over $X_a$ will yield
another gamma integrator --- in which case the integral with respect
to $\mathcal{D}c$ is well defined. Explicitly,
\begin{eqnarray}
\int_{X_a}\delta_{S_s}(c'(x))\mathcal{D} \omega_{\xx,\mathrm{Q}}(x)
&=&\int_{B_a}\mathcal{D}\omega_{\xx,\mathrm{Q}}(x)
\mathcal{D}\gamma_{1,ic'(x),\infty}(c)\notag\\
&=:&\int_{C}\big\langle \,e^{- i\langle
S_s(c'(x)),c\rangle}\big\rangle_{\omega_{\xx,\mathrm{Q}}}\mathcal{D}\gamma_{1,0,\infty}(c)
\end{eqnarray}
and the Gaussian expectation of $\exp\{- i\langle
S_s(c'(x)),c\rangle\}$ must lie in the family of gamma integrators.
Likewise, $\big\langle \exp\{-\pi/\s\,
\mathrm{Q}(x)\}\big\rangle_{\gamma_{1,ic'(x),\infty}}$ must lie in
the family of Gaussian integrators --- which brings us to the next subsection where the constraint is Gaussian-type.

\subsection{Quotient spaces with a principal bundle structure}
Let ${\M}$ be a Riemannian manifold without boundary and
$\pi_{\mathbb{G}}:\PP\rightarrow{\M}$
 a principal fiber bundle endowed with a connection. We wish to
define the  functional integral
$\int_{{M_a}}\mathrm{F}_{{\mu}}({m})
\mathcal{D}{m}$ where ${M_a}\ni
{m}:[\ti_a,\ti_b]\rightarrow{\M}$. The problem
is that the base space is complicated in general: It may be very
difficult or impossible to \emph{directly} define an integrator on
${\M}$.

On the other hand, the covering space is usually easier to handle,
and we assume that we can define an integrator so that the integral
$\int_{{P_a}}\mathrm{F}_{{\mu}}(p) \mathcal{D}p$ is well-defined. We also
assume that $\mathrm{F}_{{\mu}}(p)(\ti)$ furnishes a representation of
$\mathbb{G}$ and is equivariant so that
\begin{equation}
\mathrm{F}_{{\mu}}(p\cdot \g)(\ti) =\rho(\g^{-1})\mathrm{F}_{{\mu}}(p)(\ti)
\end{equation}
where $\rho$ is a possibly non-faithful representation of
$\mathbb{G}$.

Now, in the CDM scheme expressions like $p(\ti)$ are shorthand for a
parametrized curve: $p(b)(\ti)={\mathrm{p}_a}
\cdot{\Sigma}(\ti,{b})$ where ${b}\in {B_a}$ and ${\Sigma}(\ti,{b}):
{\PP}\rightarrow{\PP}$ is a global transformation such that
${\mathrm{p}_a}\cdot{\Sigma}(\ti_a,\cdot)={\mathrm{p}_a}$. This parametrization allows
integrals over the generally non-Banach space ${P_a}$ to be
expressed as well-defined integrals over $B_a=X_a\times C$.

The first point to make is that the parametrization for quotient spaces is gleaned from
the local structure of the bundle $\mathbb{U}_i\times\mathbb{G}$
where $\mathbb{U}_i\subset{\M}$ and a local trivialization
is given so that $p(b)(\ti)=\big({m}(b),g(b)\big)(\ti)$.
The parametrization for the first component is dictated by the
manifold structure of ${\M}$. The parametrization for the
second component is fixed by requiring parallel transport of
$p(\ti)$ --- since this will restrict paths to ${\M}$ if
that is where they start. Consider an open set
$\mathbb{U}_i\subset{\M}$, and let $A_i$ denote the local
gauge potential relative to the \emph{canonical} local section
$s_i:{\M}\rightarrow{\PP}$. The equation for parallel
transport,
\begin{equation}\label{parallel transport}
dg_i(\ti)=-g_i(\ti)A_i(\dot{{m}}(\ti))d\ti
\end{equation}
clearly indicates the conditional relation between ${m}$
and $g$, and it emphasizes the interplay between constraints and
conditionals.

To attack this problem we need to be more explicit about the
parametrization of $P_a$.
\begin{definition}
Let $\{\Bold{\omega}_i=0\}$ where $\Bold{\omega}_i\in\Lambda T^\ast
P_a$ and $i\in\{1,\ldots,r\}$ be an exterior differential system
with integral manifold $(B,P_a)$. This system defines a
parametrization $\mathcal{P}:B\rightarrow P_a$ by
\begin{equation}
\mathcal{P}^\ast\Bold{\omega}_i=0\;\forall\;i.
\end{equation}

If $B=X\times Y$ and $i=2$, the parametrization can
be written locally on $\PP$ as
\begin{equation}\label{parametrization}
\left\{\begin{array}{ll}
        d{m}(x)(\ti)=\Bold{X}_{(a)}(p(\ti))dx(\ti)^a
        \;\;\;\;\;\;
        {m}(\ti_a) = {\mathrm{m}_a}\\
        dg(y)(\ti)=\Bold{Y}_{(b)}(p(\ti))dy(\ti)^b
        \;\;\;\;\;\;\;\;
        g(\ti_a) = {\mathrm{g}_a}
      \end{array}\right.
\end{equation}
where $a\in\{1,\ldots,p_a\}$, $b\in\{1,\ldots,p_b\}$,
$p_a+p_b\leq\mathrm{dim}\,\PP$, and the set
$\{\Bold{X}_{(a)},\Bold{Y}_{(b)}\}$ generates a vector sub-bundle
$\mathbb{V}\subseteq T\PP$ of the tangent bundle. The solution of
\emph{(\ref{parametrization})} will be denoted
$p(x,y)(\ti)=\mathrm{p}_a\cdot\Sigma(\ti,x,y)$ where
$\Sigma(\ti,x,y):\PP\rightarrow\PP$ is a global transformation on the covering space such
that $\Sigma(\ti_a,\cdot,\cdot)=\mathrm{Id}$.
\end{definition}

The parallel transport equation implies $d\log(g(\ti))\sim A\cdot
d{m}(\ti)$ which implicitly encodes the constraint through
(\ref{parametrization}). This particular parametrization yields
(within $\mathbb{U}_i$)
\begin{equation}
p({x})(\ti)
=\big({m}({x}),g({m}({x}))\big)(\ti)
=\big({\mathrm{m}_a}\cdot{\sigma}(\ti,{x})
\,,\,\mathrm{g}_a\cdot\mathcal{T}
e^{-\int_{\ti_a}^{\ti}A\cdot\;\dot{{m}}\;d\ti}\big)
\end{equation}
where ${\sigma}=\pi({\Sigma})$. On the other hand, since
$p$ is a horizontal lift it can be represented as
\begin{equation}
p({x})(\ti) =s_i({m}({x})(\ti))\cdot
g({m}({x}))(\ti)\;.
\end{equation}
It is clear that the restriction expressed by (\ref{parallel
transport}) implies
$g({m}({x}))(\ti)\in\mathbb{G}\;\forall\; \{\ti,{x}\}$,
and $p(x):[\ti_a,\ti_b]\rightarrow{\PP(\mathrm{p}_a)}$ where
$\mathbb{H}_{(\mathrm{p}_a)}
\rightarrow{\PP(\mathrm{p}_a)}\stackrel{\Pi}{\rightarrow}{\M}$ is
the holonomy bundle.

We have learned that, given ${m}({x})$ and a local
trivialization, the functional constraint
$\delta(M(g))=\delta(\dot{g}-g\,A\cdot\dot{{m}})$ will
lead to a path confined to an open neighborhood of a section
isomorphic to the base space. Since the constraint only shifts
the path along fibers, we expect it to be realized by a delta functional on $G_a$ which
implies a Dirac integrator $\mathcal{D}\delta_{g_e}(M(g))$ should be
used. Moreover, $M$ just effects translation on $G_a$ so the functional
determinant of $M'$ is trivial and the zero locus coincides with the
holonomy group $\mathbb{H}_{(\mathrm{p}_a)}$.

So, for a Gaussian integrator,
\begin{eqnarray}
\int_{{M_a}}\mathrm{F}_{{\mu}}({m})\mathcal{D}\omega({m})
&:=&\int_{{X_0}} \int_{G_a}\mathrm{F}_{\mu}(\mathcal{P}(x,y)) \delta(M(g))\;
\mathcal{D}g\,\mathcal{D}\omega_{\xx,\mathrm{Q}}(\mathcal{P}({x}))
\notag\\
&=&\int_{\mathbb{H}_{(\mathrm{p}_a)}}\int_{{X_0}}
\mathrm{F}_{\mu}(\mathcal{P}({x})\cdot\mathrm{h})\;
\mathcal{D}\omega_{\xx,\mathrm{Q}}(\mathcal{P}({x}))\;d\mathrm{h}\notag\\
&=&\int_{\mathbb{H}_{(\mathrm{p}_a)}}\int_{{X_0}}
\rho(\mathrm{h}^{-1})\mathrm{F}_{\mu}(\mathcal{P}({x}))\;
\mathcal{D}\omega_{\xx,\mathrm{Q}}(\mathcal{P}({x}))\;d\mathrm{h}
\end{eqnarray}
where $\mathcal{P}({x})(\ti)\in{\PP(\mathrm{p}_a)}$.

In particular, the point-to-point propagator on ${\M}$ (in the $\rho^r$ representation)
obtains for the familiar choice $\mathrm{F}_{\mu}(\mathcal{P}({x}))(\ti_b)
=\delta(p({x})(\ti_b),\mathrm{p}_b)$;
\begin{eqnarray}
{K}^r({\m_a},{\m_b})
&:=&\int_{{M_a}}\delta({m}(\ti_b),{\m_b})\;
\mathcal{D}\omega({m})\notag\\
&=&\int_{\mathbb{H}_{(\mathrm{p}_a)}}\rho^r(\mathrm{h}^{-1})\int_{{X_0}}
\delta(p({x})(\ti_b),{\mathrm{p}_b})\;
\mathcal{D}\omega_{\xx,\mathrm{Q}}(\mathcal{P}({x}))\;d\mathrm{h}\notag\\
&=:&\int_{\mathbb{H}_{(\mathrm{p}_a)}}\rho^r(\mathrm{h}^{-1})
{K}_{\mathrm{h}_b}({\m_a},{\m_b})\; d\mathrm{h}
\end{eqnarray}
where the propagator
${K}_{\mathrm{h}_b}({\m_a},{\m_b})$ is the
point-to-point propagator on ${\PP(\mathrm{p}_a)}$ --- associated with the
homotopy class of paths indexed by ${\mathrm{h}_b}$ --- pulled back
to ${\M}$. It is legitimate to write
${K}_{\mathrm{h}_b}({\m_a},{\m_b})$ instead of
${K}({\mathrm{p}_a},\mathrm{p}_b)
={K}(s_i({\m_a}),s_i({\m_b})\cdot\mathrm{h}_b)$
because the horizontal lifting does not depend on the trivialization
and, hence, the \emph{canonical} section $s_i$. So we are
free to choose the trivial section.

In many dynamical systems of interest, the integrator
$\mathcal{D}\omega_{\xx,\mathrm{Q}}(\mathcal{P}({x}))$ is invariant under the restricted
holonomy group $\mathbb{H}_{(\mathrm{p}_a)} ^0$. Then
${K}(\mathrm{p}_a,\mathrm{p}_b\cdot\mathrm{h}^0)={K}(\mathrm{p}_a,\mathrm{p}_b)$ where
$\mathrm{h}^0\in\mathbb{H}_{(\mathrm{p}_a)} ^0$. In particular, for a one-dimensional representation, the propagator on the
base space reduces to the well-known result (for
$\chi:\mathbb{G}\rightarrow\mathbb{T}\subset\C$)
\begin{equation}
{K}^{{\Lambda_{\mathrm{p}_a}}}
({\m_a},{\m_b})
=\sum_{g\in\mathbb{G}}
\chi^{\Lambda_{\mathrm{p}_a}}(g){K}_{[\,\mathrm{h}_b]}({\m_a},{\m_b})
\end{equation}
where $\Lambda_{\mathrm{p}_a}$ labels inequivalent one-dimensional unitary representations of the monodromy
group $\mathbb{G}=\mathbb{H}_{(\mathrm{p}_a)} /\mathbb{H}_{(\mathrm{p}_a)} ^0$ at
the point $\mathrm{p}_a$ and the equivalence classes $[\,\mathrm{h}_b]$ depend on $g\in \mathbb{G}$.

Using (\ref{full propagator}),
we get what can be interpreted as a \emph{semi-classical} trace formula for point-to-point transitions on $\M$ in terms of the monodromy group
\begin{equation}
\sum_{\bar{{m}}}\frac{e^{\pi i \,\mathrm{B}(\bar{{m}}(\ti_b))}}
{\sqrt{\mathrm{det}\left[i\Bold{G}(\ti_b,\ti_b)\right]}}
=\sum_{\Lambda_{\m_a}}\sum_{g\in\mathbb{G}}d_{\Lambda_{\m_a}}
\chi^{\Lambda_{\mathrm{p}_a}}(g)\,
{K}_{[\,\mathrm{h}_b]}({\m_a},{\m_b})
\end{equation}
with $d_{\Lambda_{\m_a}}$ the multiplicity of $\Lambda_{\m_a}$.

\subsection{Bounded configuration space}\label{boundaries}
This type of system is interesting, because it requires a gamma integrator both for the non-dynamical degree of freedom $\tau:[\ti_a,\ti_b]\rightarrow\R_+$ associated with fixed boundaries and for the constraint that enforces the boundary conditions.

We wish to define an integrator for a
space of pointed paths $M_a^\p$ with $m:[\tau_a,\tau_b]\rightarrow\M$
and $\p\M\neq\emptyset$ sufficiently regular.  Experience suggests we consider a product manifold $\N=\M\times\R_+$ and take $B_a=X_a\times{\Ta}\times C$ to impose the required constraints. The task is to make sense of an integral of the form
$\int_{M_a^\p}\mathrm{F}_{{\mu}}({m})\mathcal{D}{m}$.

The parametrization of
$n:=(m,\tau):[\tau_a,\tau_b]\times[\ti_a,\ti_b]\rightarrow\N$ can be written
\begin{equation}
\left\{\begin{array}{ll}
        dm(x)(\tau)=\Bold{X}_{(a)}(m(x)(\tau))dx(\tau)^a
        \;\;\;\;\;m(x)(\tau_a) = \mathrm{m}_a
        \\
        d\tau(\ti)=\Bold{Y}(\tau(\ti))d\ti\hspace{1.2in}\tau(\ti_a)
        =\tau_a
      \end{array}\right.\;.
\end{equation}
To make the notation manageable, let $Y_a$ stand for $X_a\times{\Ta}$. Then we can write simply $y\equiv(x,\tau)$. We also put\footnote{We claim that $\mathcal{D}\gamma_{0,0,\infty}(\tau)$ is the correct integrator to use on $T_0$ because $\tau(\ti)\in\R_+$ and $\tau$ is non-dynamical.} $\mathcal{D}\Omega_{\bar{y},{\mathrm{Q}}}(y):=\mathcal{D}\omega_{\xx,{\mathrm{Q}}}(x)\,\mathcal{D}\gamma_{0,0,\infty}(\tau)$. Note that for the $\mathcal{D}\Omega$
integrator it is consistent to use a gamma integrator to account for
the boundary constraints since it is conjugate to $\mathcal{D}\Omega$.

Define the integral by\footnote{We won't motivate this definition, but the reader is invited to attach interpretations and intuitions to the various representations presented in the definition.}
\begin{eqnarray}
\int_{M_a^\p}\mathrm{F}_{{\mu}}({m})\mathcal{D}{m}
&:=&\int_{Y_a\times C}\mathrm{F}_{\mu}(n(y)|c))\Theta_{Y|C}((n(y)|c,\cdot)
\mathcal{D}_{\Theta_{Y|C},\mathrm{Z}_{Y|C}}n(y)|c\notag\\
&=:&\int_{Y_a\times C}\widetilde{\mathrm{F}}_{\mu}(S_s(n(y)),c,\cdot)
\mathcal{D}\Omega_{\bar{y},{\mathrm{Q}}}(n(y))\mathcal{D}\gamma_{\alpha,ic',\infty}(c)\notag\\
&=&\int_{{C}}\left\langle\widetilde{\mathrm{F}}_{\mu}(S_s(n(y)),c)\right\rangle_{\bar{y},{\mathrm{Q}}}
\mathcal{D}\gamma_{\alpha,ic',\infty}(c)\notag\\
&=:&\int_{{C}}\widetilde{\mathrm{H}}_{\mu}(c)
\mathcal{D}\gamma_{\alpha,ic',\infty}(c)\;.
\end{eqnarray}
It remains to infer the nature of
$\widetilde{\mathrm{F}}_{\mu}(S_s(n({y})),c)$ and the associated
integrator parameters.

From the variational principle, we learned that the constraints
impose transversality conditions on \emph{critical} paths.  We also
learned that $\tau$ should be viewed as a non-dynamical degree of freedom that reparametrizes $m(x)$. So the plan is to search for a
conditional integrator for which\footnote{The same idea was implicit
in the quotient space analysis. There, $\Theta_{X|G}$ was determined
by (\ref{parallel transport}).}
\begin{equation}
\Theta_{C|X}(c|x,\cdot)
\propto\Theta_{S_s(X)|C}\left(S_s(x),\cdot\right)
\Theta_C(c,\cdot)
\end{equation}
where $S_s(X)$ is determined by \emph{mean} paths. We replace
critical with mean paths for the sufficient statistic because the
quantum analog of the variational principle is
\begin{equation}
\frac{\delta\Gamma(x)}{\delta x(\ti)}\sim x'(\ti)
\end{equation}
with $\Gamma$ defined in eq. (B.10) \cite{LA4}.  Furthermore, we restrict to the case $\tau(\ti)$ real.

Having identified the relevant sufficient statistics for the paths, we now switch solution strategies and work instead with $\widetilde{\mathrm{G}}_{\mu}(S_s(c),n(y))$. To simplify matters, only the two limiting cases of transversal
intersection and fixed energy discussed in \S 2 \cite{LA4}
 will be considered. Recall that these
cases correspond to point-to-boundary and point-to-point paths
respectively. Let $\bar{n}=(\bar{m},\bar{\tau})$ represent a mean
path. Define the ``mean exit time'' $\tau_o$ implicitly by $\bar{n}(\ti_b)=(\bar{m}(\tau_o),\bar{\tau}(\ti_b))$ such that $\bar{m}(\tau_o)\in\p\M$ where $\overline{\tau}(\ti_b):=\tau_o$ characterizes the mean gamma process. Recall
the mean path $\bar{n}$ is also critical since $\mathrm{Q}$ is quadratic, and
there may be more than one critical path.

The functional integral for a functional $\mathrm{F}^\p_\mu$ that takes its
values on $\p\M$ is defined by
\begin{eqnarray}
\Phi^\p(\m_a)&:=& \int_{Y_a}\langle\widetilde{\mathrm{F}}^\p_{\mu}\left(m(x)\right)\rangle_{\gamma_{1,ic'(\tau),\infty}}
\;\mathcal{D}\Omega_{\bar{y},{\mathrm{Q}}}(n(y))
\end{eqnarray}
where
\begin{equation}
\langle\widetilde{\mathrm{F}}^\p_{\mu}\left(m(x)\right)\rangle_{\gamma_{1,ic'(\tau),\infty}}
:=\int_{T_0\times C}\widetilde{\mathrm{F}}^\p_{\mu}\left(m(x)(\tau(\ti_b))\right)\,\mathcal{D}\gamma_{1,Id',\infty}(\tau)
\,\mathcal{D}\gamma_{1,ic'(\tau),\infty}(c)
\end{equation}
with $i\langle S_s(c'(\tau)),c\rangle=-2\pi i[\tau(\ti_b)-\bar{\tau}(\ti_b)]\cdot\bar{c}$ and $\overline{\tau}(\ti_b)=\tau_o$.

As we have seen before, the constraint is just a delta functional on $C'$ and the integral reduces; (restoring $y\rightarrow (x,\tau)$ for clarity)
\begin{eqnarray}
\Phi^\p(\m_a)&=&\int_{X_a\times T_0}\langle\widetilde{\mathrm{F}}^\p_{\mu}\left(m(x)\right)\rangle_{\gamma_{1,ic'(\tau),\infty}}
\,\mathcal{D}\omega_{\bar{m},\widetilde{\mathrm{Q}}}(x)\,\mathcal{D}\gamma_{0,0,\infty}(\tau)\notag\\
&=&\int_{X_a\times\R_+}\widetilde{\mathrm{F}}^\p_{\mu}\left(m(x)(\tau_o)\right)
\,\mathcal{D}\omega_{\bar{m},\widetilde{\mathrm{Q}}}(x)\,d(ln(\tau_o))\notag\\
&=&\int_{\R_+}\langle\widetilde{\mathrm{F}}^\p_{\mu}\left(m(x)(\tau_o)\right)\rangle_{\bar{m},\widetilde{\mathrm{Q}}}\;d(ln(\tau_o))
\end{eqnarray}
where $\widetilde{\mathrm{Q}}:=\mathrm{Q}\circ m$.

Similarly, for a functional $\mathrm{F}_\mu^{\backslash\p}$ that takes its
values in $\M \backslash\p\M$,
\begin{eqnarray}
\Phi^{\backslash\p}(\m_a):=\int_{Y_a}
\langle\widetilde{\mathrm{F}}^{\backslash\p}_{\mu}\left(m(x)\right)\rangle_{\gamma_{0,ic'(\tau),\infty}}
\;\mathcal{D}\Omega_{\bar{y},{\mathrm{Q}}}(n(y))
\end{eqnarray}
where
\begin{eqnarray}
\langle\widetilde{\mathrm{F}}^{\backslash\p}_{\mu}\left(m(x)\right)\rangle_{\gamma_{0,ic'(\tau),\infty}}
&:=&\int_{T_0\times C}\widetilde{\mathrm{F}}^{\backslash\p}_{\mu}\left(m(x)(\tau_b)\right)\,\mathcal{D}\gamma_{1,Id',\infty}(\tau)
\,\mathcal{D}\gamma_{0,ic'(\tau),\infty}(c)\notag\\
&=&\int_{\R_+}\theta(\tau_o-\tau_b)\widetilde{\mathrm{F}}^{\backslash\p}_{\mu}\left(m(x)(\tau_b)\right)\,d\tau_b
\end{eqnarray}
Note the step functional constraint in this case.

These definitions also hold for
$\mathrm{Q}\rightarrow \mathrm{S}$, but the mean paths are no longer critical. Of
course, the suitability of these definitions rests on their ability
to reproduce known results (see \cite{LA2} for some examples).

Like the quotient space example, the two integrals simplify for propagators. For the point-to-boundary
propagator\footnote{The need for the normalization constant
$\mathcal{N}(\m_a)$ can be established from dimensional analysis.},
\begin{eqnarray}
K_\p(\m_a,\m_B)&=&\mathcal{N}(\m_a)\int_{Y_a}
\,\delta(m(x)(\tau_o),\m_B)
\;\mathcal{D}\Omega_{\bar{y},{\mathrm{Q}}}(n(y))
\notag\\
&=&\mathcal{N}(\m_a)\sum_{\bar{m}(\tau_o)}\int_{\R_+}\frac{e^{\pi i
\mathrm{B}(\bar{m}(\tau_o))}}{\sqrt{\mathrm{det}i\Bold{G}(\tau_o,\tau_o)}}
\;d(\ln\tau_o)
  \end{eqnarray}
where $\m_B\in\p\M$ denotes an end-point on the boundary and the normalization
$\mathcal{N}(\m_a)$ enforces $\int_{\p}K_\p(\m_a,\m_B)d\m_B=1$. For example, if $\M\subset\R^n$, then
the boundary term goes like $\mathrm{B}(\bar{m}(\tau_o))\sim
|\m_B-\m_a|^2/\tau_o$. If there is more than one critical (or
mean) path, care must be taken to split the integral over the
boundary into regions associated with a particular critical path.

For the point-to-point propagator, it is convenient to first integrate with respect to $\mathcal{D}\omega$;
\begin{eqnarray}\label{p-to-p}
K(\m_a,\m_b)&=&\int_{Y_a\times C}
\,\langle\delta(m(x),\m_b)\rangle_{\gamma_{0,ic'(\tau),\infty}}
\;\mathcal{D}\Omega_{\bar{y},{\mathrm{Q}}}(n(y))\notag\\
&=&\sum_{\bar{m}^E}\int_{T_0}\int_{\R_+}\theta(\tau_o-\tau_b)\frac{e^{\pi i
\mathrm{B}(\bar{m}^E)}}
{\sqrt{\mathrm{det}i\Bold{G}(\tau_b,\tau_b)}}\;d\tau_b\;\mathcal{D}\gamma_{0,0,\infty}(\tau)
\end{eqnarray}
where $\m_b:=m(\tau_b)$. The step function makes this propagator fairly difficult to handle since $\tau_o$ implicitly depends on $\m_a$. One way around the difficulty is to write the inner integral as
\begin{equation}
\int_{\R_+}\theta(\tau_o-\tau_b)\frac{e^{\pi i
\mathrm{B}(\bar{m}^E)}}
{\sqrt{\mathrm{det}i\Bold{G}(\tau_b,\tau_b)}}\;d\tau_b=\left(\int_{\R_+}-\int_{\tau_o}^\infty\right)\frac{e^{\pi i
\mathrm{B}(\bar{m}^E)}}
{\sqrt{\mathrm{det}i\Bold{G}(\tau_b,\tau_b)}}\;d\tau_b
\end{equation}
and then find a point transformation on $\M$ that takes $\m_a$ to the boundary (see \cite{LA2}).\footnote{The point transformation then implies $\tau_o\rightarrow0$ and $\bar{m}^E$ is transformed accordingly in the second integral on the right. Since there is no longer any $\tau_o$ dependence in this case, the outer integral in (\ref{p-to-p}) just contributes its normalization (which we have set by the choice $\int_{T_0}\mathcal{D}\tau=1$). The same thing happens for the unbounded case when $\tau_o\rightarrow\infty$ in the step function.}

These point-to-point and point-to-boundary propagators are \emph{restricted} in the sense that the paths
are not allowed to penetrate the boundary. However, there are cases
of interest when the boundary represents a discontinuity, and the
paths are defined on both sides of the boundary.

\subsection{Segmented configuration space} Often the target space of paths
will have codimension-1 submanifolds that induce a decomposition of
path space $\bigoplus _iX_{a}^{(i)}=X_a$ in the sense that each
$X_{a}^{(i)}$ has its own integrator. Such is the case, for example,
when a Gaussian integrator is defined in terms of an action
functional and the potential in the action is discontinuous. More
explicitly, $x_{a}^{(i)}\in X_{a}^{(i)}$ is the pointed path
$x_{a}^{(i)}:[\ti_a,\ti_b]\rightarrow(\M^{(i)},\m_{a}^{(i)})$ where
$\M=\bigcup_i\M^{(i)}$ such that the intersection
$\M^{(i)}\cap\M^{(j)}=\p\M^{(ij)}$ is a submanifold of
codimension-1.

The objects of interest in this case are the propagators from the
previous subsection. Since it is known that the propagators are
kernels for certain differential operators, Green's theorem provides
a convenient starting point for the analysis.

For an operator $L$ defined in a bounded open region $\U\subset\M$
acting on complex scalar functions from an appropriate function
space,
\begin{equation}
\int_{\U}(L\phi)\overline{\varphi}
-\int_{\U}\phi\overline{(L^{\ast}\varphi)}
=\int_{\p\U}B(\phi,\overline{\varphi})
\end{equation}
where $L^\ast$ is the formal adjoint of $L$, and the functional form
of $B$ is determined from Stoke's theorem and the particular
boundary conditions associated with the function space.

In particular, let $\U_1=\U^{(1)}\cup\U^{(2)}$ be a bounded open
region in $\R^3$ with one surface $\mathbb{S}=\U^{(1)}\cap\U^{(2)}$
of discontinuity. Choose $\varphi$ to be the Green's function of
$\nabla^{\ast}$ in $\U_1$ with vanishing Dirichlet boundary
conditions on $\p\U_1$ and $\phi$ the Green's function of $\nabla$
in $\U^{(1)}$ with vanishing Dirichlet conditions on $\mathbb{S}$.
Then the theorem gives the Green's function $\overline{\varphi}$ of
$\nabla$ in $\U^{(1)}$ with \emph{non-vanishing} boundary conditions
on $\mathbb{S}$
\begin{equation}
\overline{\varphi}=\phi+\int_{\mathbb{S}}\overline{\varphi}\,\nabla\phi\
\cdot {d}\sigma
\end{equation}
and in $\U_1\setminus\U^{(1)}$
\begin{equation}
\overline{\varphi}=\int_{\mathbb{S}}\overline{\varphi}\,\nabla\phi\
\cdot {d}\sigma\;.
\end{equation}

This theorem has a simple interpretation in terms of functional
integral representations of propagators: For bounded regions that
allow paths to penetrate the boundary, the total point-to-point propagator includes the
\emph{restricted} point-to-point propagator (which does not allow
paths to leave the region) plus the potential on the bounding
surface induced by all sources accessible to paths that are allowed
to leave the region. This prescription is equivalent to the ``path
decomposition technique'' used in \cite{AU/SC}--\cite{HA}.

In other words, Green's theorem can be used to partition the space
of paths taking their values in a segmented configuration space into
restricted and unrestricted sets. This is useful because the paths
in the partitioned sets have particularly convenient boundary
conditions and their associated propagators are relatively easy to
calculate.

Let $\K_{\U^{(i)}}^{(D)}$ be the restricted point-to-point
propagators with Dirichlet boundary conditions on $\mathbb{S}$, and
$\K_{\p^{(i)}}^{(D)}$ the restricted point-to-boundary homogenous
propagators for their respective regions $\U^{(i)}$. These are the
propagators derived in the previous subsection. According to Green's
theorem, the \emph{unrestricted} point-to-point propagator from
$\m_a^{(i)}$ to $\m_b^{(j)}$ in $\U_1$ with one surface of
discontinuity can be written
\begin{equation}\label{split kernel}
  \K_{\U_1}^{(D)}(\m_a^{(i)},\m_{b}^{(j)}):=
  \delta_{ij}\K_{\U^{(j)}}^{(D)}(\m_a^{(i)},\m_{b}^{(j)})
  +\int_{\mathbb{S}}\K_{\p^{(i)}}^{(D)}(\m_a^{(i)},\sigma)
  \K_{\U_1}^{(D)}(\sigma,\m_{b}^{(j)})\;d\sigma
\end{equation}
where $\m^{(i)}\in\U^{(i)}$ and $\sigma\in\mathbb{S}$. Intuitively,
the unrestricted point-to-point propagator within a bounded region
$\U^{(i)}$ is implicitly determined by the restricted point-to-point
and point-to-boundary propagators in $\U^{(i)}$. Similarly, the
point-to-point propagator from $\U^{(i)}$ to $\U^{(j)}$ is
implicitly determined by the restricted point-to-boundary propagator
in $\U^{(i)}$. Note that $\K_{\U_1}^{(D)}$ has non-trivial boundary
conditions on $\mathbb{S}$, but it still satisfies Dirichlet
boundary conditions on $\p\U_1$.

Equation (\ref{split kernel}) is a familiar expression, and it is
often solved by iteration. However, the path
decomposition idea along with the fact that $\K_{\U_1}^{(D)}$
\emph{inside the integral is evaluated on the surface},
suggests\footnote{This argument is admittedly highly heuristic.}
that we replace ${\K}_{\U_1}^{(D)}$ (inside the integral) with
$\widetilde{\K}_{\U_0}^{(D)}$ defined by
\begin{equation}
\widetilde{\K}_{\U_0}^{(D)}(\sigma,\m_b^{(j)}):=\left\{\begin{array}{cc}
r_{(\neg i)}(\sigma)\K_{\U_0|_i}^{(D)}(\m_{a'}^{(i)},\m_{b}^{(i)})
|_{\m_{a'}^{(i)}=\sigma} \; \mathrm{if}\;i=j \\
t_{(\neg j)}(\sigma)\K_{\U_0|_j}^{(D)}(\m_{a'}^{(j)},\m_{b}^{(j)})
|_{\m_{a'}^{(j)}=\sigma} \; \mathrm{if}\;i\neq j
                 \end{array}\right.\;,
\end{equation}
where $\K_{\U_0|_j}^{(D)}$ is the unrestricted propagator evaluated
in $\U^{(j)}$, and
\begin{eqnarray}
r_{(\neg i)}(\sigma)&=&\int_{\mathbb{S}}
  \K_{\U_0|_{\neg i}}^{(D)}(\sigma,\sigma')\K_{\p^{(\neg
i)}}^{(D)}(\sigma',\sigma)\;d\sigma'\notag\\
t_{(\neg j)}(\sigma)&=&\int_{\mathbb{S}}
  \K_{\U_0|_{\neg j}}^{(D)}(\sigma,\sigma')\K_{\p^{(\neg
j)}}^{(D)}(\sigma',\sigma)\;d\sigma'\;.
\end{eqnarray}
are evaluated in the region on the other side ($\neg j$) of
$\m_b^{(j)}$. According to the path decomposition picture,
$r(\sigma)$ and $t(\sigma)$ measure the (probability amplitude)
contribution of pointed loops based at $\sigma$ that lie on either
side of $\mathbb{S}$ and so $|r|^2+|t|^2=1$.

Usually it is simpler to find $\widetilde{\K}_{\U_0}^{(D)}$
than to iterate (\ref{split kernel}). For example, consider the standard elementary text book example of the fixed
energy propagator in $\R$ with a step potential $V(x)=V_0\theta(x)$.
Let $k_0$ and $k_{V_0}$ be the wave vectors to the left and right of
$x=0$ respectively. Then it is easy to see from the surface integral
that $t\sim 2\sqrt{k_0k_{V_0}}/(k_0+k_{V_0})$ since this must be
symmetric under $k_0\leftrightarrow k_{V_0}$; and, hence, $r\sim
(k_0-k_{V_0})/(k_0+k_{V_0})$ from the normalization condition.

Now let $\U$ be divided into three regions. We can use the results
for a single surface of discontinuity by covering $\U$ with two
overlapping copies of $\U_1$. That is, each region contains only one
surface of discontinuity. There are now six relevant propagators
with appropriate boundary conditions. Their nature depends on
whether or not the boundaries intersect $\p\U$. Since we require
Dirichlet boundary conditions for point-to-point transitions and
$\K_{\U_1}^{(D)}$ can only propagate across a single discontinuity,
care must be taken to use the appropriate $\U_1$ for any given
transition.

To simplify, specialize to a planar geometry, and order the regions
$\{1,2,3\}$. There will be two classes of propagators;
half-space-type in regions 1 and 3 defined by
$\overline{\U_1\cap\U_{1'}}$, and unit-strip-type in region 2
defined by $\U_1\cap\U_{1'}$ where $\U_1=\U^{(1)}\cup\U^{(2)}$ and
$\U_{1'}=\U^{(2)}\cup\U^{(3)}$. Choose the partition
$\U=\U_1\cup\U^{(i)}$ so that $\U_1\equiv\U^{(j)}$ contains the
initial point $x_{a'}^{(j)}$. By combining the single surface result
appropriately, the \emph{approximate} point-to-point propagator for two surfaces of
discontinuity can be written
\begin{eqnarray}\label{3 region}
  \K_{\U_{2}}^{(D)}(\x_a^{(i)},\x_{a'}^{(j)})&\simeq&
  \delta_{ij}\K_{\U^{(j)}}^{(D)}(\x_a^{(i)},\x_{a'}^{(j)})\notag\\
  &&+\int_{\mathbb{S}_2}
  \K_{\p^{(j)}}^{(D)}(\x_a^{(i)},\sigma^{(j)})
  \widetilde{\K}_{\U_1}^{(D)}(\sigma^{(j)},\x_{a'}^{(j)})\;d\sigma^{(j)}\notag\\
\end{eqnarray}
where $\K_{\U^{(j)}}^{(D)}$ is the restricted point-to-point
propagator and $\K_{\p^{(j)}}^{(D)}$ the restricted
point-to-boundary propagator derived from (\ref{split kernel}).

It is important to remember that the propagator depends on
\emph{all} critical paths. For regions bounded by two planes, there
is obviously a sum over all `bounces' within the bounded
region\footnote{To the extent that the boundaries are exactly
parallel and/or the planes extend to infinity, this gives an
infinite sum which can be written analytically in the usual way as
an inverse propagator.}. These bounce transitions are encoded in the
$\K_{\p^{(j)}}(\x_a^{(i)},\sigma^{(j)})$ propagator. Hence, poles of
the convolution of relevant bounce propagators yield spectral
information for their corresponding regions.

In general then, the \emph{approximate} point-to-point propagator for
$\U=\bigcup_{i=1}^{n+1} \U^{(i)}$ is determined recursively from
$\K_{\U_{0}}^{(D)}$, which is the standard unrestricted
point-to-point elementary kernel in the region $\U^{(j)}$ with
vanishing Dirichlet boundary conditions on $\p\U$, and
\begin{eqnarray}\label{recursion}
  \K_{\U_{n}}^{(D)}(\x_a^{(i)},\x_{a'}^{(j)})&\simeq&
  \delta_{ij}\K_{\U^{(j)}}^{(D)}(\x_a^{(i)},\x_{a'}^{(j)})\notag\\
  &&+\int_{\mathbb{S}_{(n)}}\K_{\p^{(j)}}^{(D)}
  (\x_a^{(i)},\sigma^{(j)})
  \widetilde{\K}_{\U_{n-1}}^{(D)}(\sigma^{(j)},\x_{a'}^{(j)})\;d\sigma^{(j)}
  \notag\\
\end{eqnarray}
where the region $\U_{n-1}$ is chosen to contain the initial point. This represents an alternative to the iteration approximation that, in particular, may be useful in numerical applications.

There are special cases when the iteration of (\ref{split kernel}) can be
summed explicitly. The method (see e.g. \cite{CR}) essentially boils
down to finding a Poisson integrator that is valid everywhere in
$X_a$. To see this, let $\mathrm{S}(x)=\mathrm{Q}(x)+\mathrm{V}(x)$ describe some general
(action) functional, and suppose the kernel $K_\mathrm{Q}$ for the time-independent Schr\"{o}dinger operator has been found
in each $\U^{(i)}$. Define $\mathrm{V}_\mathrm{Q}=K_\mathrm{Q}\circ \mathrm{V}\circ K_\mathrm{Q}$ by
its evaluation on ordered graphs in $\M$, i.e. under the linear maps
$L_n:T_0\rightarrow {i\R^n}$ we have $m(x)(\tau)\mapsto
(m(x(\tau_1)),\ldots,m(x(\tau_n)))=:(\m_1,\ldots,\m_n)\in\M^n$. Then
\begin{equation}
\mathrm{V}_\mathrm{Q}(m(x(\tau)))\rightarrow K_\mathrm{Q}(\m_b,\m_k)\mathrm{V}(\m_k)
K_\mathrm{Q}(\m_k,\m_{k-1}),\ldots,\mathrm{V}
(\m_1) K_\mathrm{Q}(\m_1,\m_a)
\end{equation}
where $\m_a \leq\m_1<\ldots\leq\m_b$ are the time-ordered graph
nodes. A time-slicing analysis \cite{CR} when $\M=\R^m$ shows the
kernel has the form
\begin{equation}\label{Ks}
K_\mathrm{S}(\m_a,\m_b)=\langle\m_a|\langle
\mathrm{V}_\mathrm{Q}\rangle_{(\ti_b-\ti_a)}|\m_b\rangle=:\langle\m_a|K_\mathrm{S}((\ti_b-\ti_a))|\m_b\rangle
\end{equation}
where $\langle \cdot\rangle_{c}$ is the Poisson expectation defined in
subsection B.0.4 \cite{LA4}. But the time-slicing analysis is only straightforward in $\R^m$;
otherwise there are well-known pitfalls.

However, the target
manifold independence of the right-hand side of (\ref{Ks}) suggests
that the definition is correct for any manifold $\M$. The point is, the right-hand side
is more  general than a perturbation expansion: It is defined in the
function space rather than on the target manifold so it offers potentially new insight and calculational techniques. Consequently, an obvious proposal is to represent
$K_\mathrm{S}(c)$ by
\begin{equation}
K_\mathrm{S}(c)\sim\int_{\mathcal{C}}\left[\int_{{\Ta}}
\mathcal{D}\gamma_{\alpha,\mathrm{S},
c}(\tau)\right]\;d\alpha
\end{equation}
for an appropriate contour $\mathcal{C}\subset\C$. This is a
substantial generalization of the Poisson expectation because now the
target space is $\C_+$ rather than $i\R$. Not only do we get phase
information not carried by the Poisson integrators, but we don't
have to restrict to
$\langle\beta',\tau\rangle=\lambda\langle{Id}',\tau\rangle$.

Here again is the thematic idea connecting propagators to summation
over an integrator family. According to the proposal, the analytic structure of
the propagator could be encoded in the integral defined by
$K_\mathrm{S}(c;\alpha):=\int_{{\Ta}}
\mathcal{D}\gamma_{\alpha,\mathrm{S}, c}(\tau)$, and $K_\mathrm{S}(\m_a,\m_b)$ would be determined by the residues of $K_\mathrm{S}(c;\alpha)$.\footnote{Of course, $K_\mathrm{S}(\m_a,\m_b)$ generally will also have singularities associated with its spectrum.}

\section{Prime examples}
So far we have used the proposed formalism to re-derive more-or-less known results using new tools. The goal in this section is to derive something new; functional integral
representations of the \emph{average} prime counting function and the average twin prime counting function. We formulate the counting functions in the spirit of
quantum mechanical expectation values in the sense that they will
represent the sum over `paths' with certain attributes.
Specifically, \emph{the paths are conjectured to follow gamma rather than Gaussian
statistics}, and restricting to prime events/numbers imposes a non-homogeneous scaling factor.

\subsection{Prime counting}
It is useful to have a physical picture in mind. Consider  a quantum system of two-state (integer/not-integer) `entities' on the positive-definite reals $\R_+$. The observables of interest are projections onto either of the two possible states.  Observation at a random point via a projector gives integer or not integer. Once a starting point and metric have been established, one knows precisely how to construct the projection operator and therefore where to observe the integers. Enumeration of the integer observations within an interval $(0,\mathsf{x}]$ then gives a correspondence between a subset of the natural numbers $\N_+$ and integer states on the lattice $\mathbb{Z}_+$ contained within the interval. This correspondence can be used to characterize/label an integer eigenstate located on $\R_+$ by its associated natural number; thus yielding a model of $\mathbb{Z}_+$ in terms of $\N_+$ as the cut-off $\mathsf{x}\rightarrow\infty$. The same goes for $\mathbb{Z}_-$. We show below that the projector onto integers follows a trivial gamma distribution, and enumeration of the integers is given by a certain trace of the associated propagator over integer states labeled by $n\in\N_+$.

To proceed, let's calculate the expected number of integers occurring up to some cut-off integer $ \mathsf{x}\in\R_+$  by defining a suitable $\alpha$-trace applied to the simple case of an homogenous process. That is, we take $\beta'=Id'$ in the lower gamma integral and impose the sufficient statistic associated with a cut-off by restricting the domain of  paths via the linear map $L:\tau\rightarrow\tau(\ti_b)$.  In this case, the functional integral can be explicitly evaluated and we get
\begin{eqnarray}\label{counting}
N( \mathsf{x})
:=\mathrm{tr}_\alpha\int_{T_0}
(-1)^\alpha\;\mathcal{D}\gamma_{\alpha,Id', \mathsf{x}}(\tau)&:=&\int_{\mathcal{C}}\frac{\Gamma(1-\alpha)}{2\pi
i}\,(-1)^{\alpha}\gamma(\alpha, \mathsf{x})\,d\alpha\notag\\
&=&\sum_{n=1}^\infty\frac{(-1)^{2n}}{(n-1)!}\,\gamma(n, \mathsf{x})\notag\\
&=&\sum_{n=1}^\infty P(n, \mathsf{x})\notag\\
&=&\Gamma(1,-\log( \mathsf{x}))= \mathsf{x}
\end{eqnarray}
where the contour encircles the positive real axis. The result $N( \mathsf{x})=\mathsf{x}$ lends credence to the choice of $\mathrm{tr}_\alpha$.

Now consider  a quantum system of two-state (prime-power/not prime-power) `entities' localized on the lattice of positive integers $\mathbb{Z}_+$. Counting `prime-power events' is postulated to be a constrained dynamical random process. As in the case of integers, we use a quantum model on $\mathbb{Z}_+$ given a suitable projector. Unfortunately, in this case we have no metric to tell where the next prime-power event will occur: Having localized onto some integer, we then must test its natural-number label for non-trivial divisors. Evidently then, states on $\mathbb{Z}_+$ posses two degrees of freedom (now including also divisors/not divisors).  This is where probability enters the process. One could interpret the integer counting as a `classical' process while counting prime powers would correspond to a random `quantum' process.\footnote{Fine, but once the prime powers are located in any given interval how can we say they are randomly distributed? Well, as long as our hypothetical quantum system that models the interval is a closed system, the observed state eigenvalues and natural-number labels remain valid and the system is deterministic. This is consistent with the persistence of quantum-state eigenvalues in the absence of external interactions. But if an external agent were to act on the system, for example by some unknown re-assignment of ordinals or re-arrangement of the points on $\mathbb{Z}_+$, then we would no longer have a correspondence between $\N_+$ and $\mathbb{Z}_+$ and the location of prime powers in the interval would have to be re-established.}

We postulate that the number of prime powers corresponds to the expectation of a suitable \emph{non-trivial} evolution operator generated by the projection onto integers --- i.e. a propagator that follows a non-trivial gamma distribution that can be represented as a constrained gamma functional integral. Accordingly, the prime counting function is the expectation of a
gamma process with \emph{unknown scale parameter} due to the constraint
associated with counting only primes. The functional integral that
enforces the constraint must be a gamma integral because the conjugate
prior of a gamma distribution with unknown scaling parameter is
again a gamma distribution. Therefore, according to the general construction, the constrained functional
integral that represents the expectation value can be written as a
constrained functional that is integrable with respect to two
marginal gamma integrators. Analogous to the QM point-to-point free
propagator example, we put $\langle
c'(\tau),c\rangle=\langle\left(\tau-\lambda(\tau)\right),c\rangle$. Then let us define the average number of primes up to some cut-off integer $ \mathrm{\mathsf{x}}$ to
be
\begin{eqnarray}
\overline{\pi_1( \mathrm{\mathsf{x}})}
&:=&\mathrm{tr}_\alpha\int_{\widetilde{T_0}}(-1)^\alpha\;
\mathcal{D}\gamma_{\alpha,Id',\widetilde{c}(\mathrm{\mathsf{x}})}(\widetilde{\tau})\notag\\
&=&\mathrm{tr}_\alpha\int_{{\Ta}\times C}(-1)^\alpha\;
\mathcal{D}\gamma_{\alpha,Id', \mathrm{\mathsf{x}}}(\tau)
\;\mathcal{D}\gamma_{1,ic'(\tau),\infty}(c)\notag\\
&=&\mathrm{tr}_\alpha\int_{{\Ta}}(-1)^\alpha\;
\mathcal{D}\gamma_{\alpha,Id',\lambda( \mathrm{\mathsf{x}})}({\tau})
\end{eqnarray}
where $\langle
S_s(c'(\tau)),c\rangle=\left( \mathrm{\mathsf{x}}-\lambda( \mathrm{\mathsf{x}})\right)\cdot
\overline{c}\,$  such that $\lambda( \mathrm{\mathsf{x}})$ represents an unknown
possibly non-homogenous scaling factor, and the $\alpha$-trace is defined below.\footnote{Contrary to what was done in $\S$ \ref{boundaries}, here we do not integrate over $ \mathrm{\mathsf{x}}$ because it is obviously fixed.} Note that the constraint imposes the non-homogeneous scaling factor on the cut-off.

We are counting primes so the counting should begin with the second event (since primes start with $p=2$).  Recalling the $\alpha$-trace from the counting-of-integers exercise, we propose
\begin{eqnarray}
\overline{\pi_1( \mathrm{\mathsf{x}})}&=&\mathrm{tr}_\alpha\int_{{\Ta}}(-1)^\alpha\;
\mathcal{D}\gamma_{\alpha,Id',\lambda( \mathrm{\mathsf{x}})}(\tau)
=\mathrm{tr}_\alpha\left[(-1)^{\alpha}\gamma(\alpha,\lambda( \mathrm{\mathsf{x}}))\right]\notag\\
&:=&\frac{1}{2\pi i}\int_{\mathcal{C}_{+1}}\frac{\pi\csc(\pi(\alpha+1))}{\Gamma(\alpha+1)}
(-1)^{\alpha}\gamma(\alpha,\lambda( \mathrm{\mathsf{x}}))\;d\alpha\notag\\
&=&\sum_{n=1}^\infty\frac{(-1)^{2n+1}}{n!}\,\gamma(n,\lambda( \mathrm{\mathsf{x}}))\notag\\
&=&-\sum_{n=1}^\infty\frac{\Gamma(n)}{\Gamma(n+1)}\,P(n,\lambda( \mathrm{\mathsf{x}}))
\end{eqnarray}
where the new contour begins at $\infty$ above the real axis, circles
the point $\{1\}$ counter-clockwise, and returns to $\infty$ below the real
axis. Roughly speaking, this calculation simply
sums the positive integers appropriately adjusted with a
non-homogenous scaling factor and weighted by
$\Gamma(n)/\Gamma(n+1)=1/n$ (which motivated the choice of $\mathrm{tr}_\alpha$).

The series converges absolutely since
\begin{equation}
\lim_{n\rightarrow\infty} \left|\frac{n!}{(n+1)!}
\frac{\left|\gamma(n+1,\lambda( \mathrm{\mathsf{x}}))\right|}
{\left|\gamma(n,\lambda( \mathrm{\mathsf{x}}))\right|}\right|
=\lim_{n\rightarrow\infty} \left|\frac{1}{(n+1)}\right|
\lambda( \mathrm{\mathsf{x}})=0\;.
\end{equation}
And notice that
\begin{equation}
\overline{\pi_1( \mathrm{\mathsf{x}}+1)}-\overline{\pi_1( \mathrm{\mathsf{x}})}=
-\sum_{n=1}^\infty\frac{1}{n!}
\int_{\lambda( \mathrm{\mathsf{x}})}^{\lambda( \mathrm{\mathsf{x}}+1)}e^{-t}\,t^{n-1}\;dt
\sim\frac{-1}{\lambda( \mathrm{\mathsf{x}})}
\end{equation}
is supposed to represent the average density of primes at $ \mathrm{\mathsf{x}}$. Accordingly, a good and obvious choice for the scaling factor is
$\lambda( \mathrm{\mathsf{x}})=-\log( \mathrm{\mathsf{x}})$ yielding the hypothesis
\begin{equation}
\overline{\pi_1( \mathrm{\mathsf{x}})}=-\sum_{n=1}^\infty\frac{\Gamma(n)}{\Gamma(n+1)}\,P(n,-\log( \mathrm{\mathsf{x}}))\;.
\end{equation}

It turns out the Moebius inversion of $\overline{{\pi}_1( \mathrm{\mathsf{x}})}$ gives essentially the same estimate as Riemann's R-function.  This can be seen by noting that  the absolute convergence of $\sum\gamma(n,-\log( \x))/n!$ together with $\int_1^\x|d\gamma(n,-\log(x))/dx|\,dx=\gamma(n,-\log(\x))$ implies
$\overline{\pi_1( \mathrm{\mathsf{x}})}=\mathrm{li}( \x)-\log(\log( \x))$. Hence $P(n,-\log( \mathrm{\mathsf{x}}))\simeq\pi( \mathrm{\mathsf{x}}^{1/n})$ and therefore
\begin{equation}\label{prime}
\overline{\pi_1( \mathrm{\mathsf{x}})}=-\sum_{n=1}^\infty\frac{1}{n}\,P(n,-\log( \mathrm{\mathsf{x}}))
 \simeq \sum_{p^k\leq \mathrm{\mathsf{x}}}\frac{1}{k}
\end{equation}
where $p^k$ is a prime power. The reader is invited to examine (\ref{prime}) numerically.

\subsection{Twin prime counting}
The constrained gamma process postulate can be applied to twin prime counting as
well. Our reasoning remains heuristic.

We maintain the hypothesis that the occurrence of twin prime numbers is a constrained  gamma process.
But now, taking pairs whose difference is $n=2$ will incur the twin
prime constant $C_2$ normalization according to the standard
probabilistic argument. Also, events between twin primes should be excluded from the $\alpha$-trace so we should only count every other event. The proposed twin prime integrator is
\begin{equation}
\frac{C_2}{\Gamma(\alpha+1)}(-1)^{2\alpha-1}\mathcal{D}\gamma_{2\alpha-1, Id',\lambda( \mathrm{\mathsf{x}})}(\tau)\;.
\end{equation}
Following the example of single primes suggests
\begin{eqnarray}
\overline{\pi_2( \mathrm{\mathsf{x}})}
&:=&\frac{C_2}{2\pi i}\int_{\mathcal{C}_{+1}}\frac{\pi\csc(\pi(\alpha+1))}{\Gamma(\alpha+1)^2}
(-1)^{2\alpha-1}\gamma(2\alpha-1,\lambda( \mathrm{\mathsf{x}}))\;d\alpha\notag\\
&=&C_2\,\sum_{n=1}^\infty\frac{(-1)^{n}}{(n!)^2}\,\gamma(2n-1,\lambda( \mathrm{\mathsf{x}}))\;.
\end{eqnarray}

Continuing this heuristic for arbitrary prime doubles
$\overline{\pi_{2i}( \mathrm{\mathsf{x}})}$ within an interval $2i\leq  \mathrm{\mathsf{x}}-2$,
it is known that the normalizing constant becomes
\begin{equation}
C_{2i}=C_2\prod_{p|i}\frac{p-1}{p-2}
\end{equation}
for prime numbers $p>2$, and this is the only adjustment to the joint integrator.  In general then, the prime double
hypothesis is
\begin{eqnarray}
\overline{\pi_{2i}({ \mathrm{\mathsf{x}}})}=C_{2i}\,\sum_{n=1}^\infty\frac{(-1)^{n}}{(n!)^2}\,\gamma(2n-1,-\log( \mathrm{\mathsf{x}}))\;,
\;\;\;\;\; \mathrm{\mathsf{x}}-2>2i\in\mathbb{N}_+\;.
\end{eqnarray}
The reader is invited to compare this average against tabulated twin primes.

Note that only the normalizing constant depends on $i$, and the
series  converges absolutely for finite $ \mathrm{\mathsf{x}}$;
\begin{equation}
\lim_{n\rightarrow\infty}
\left|\frac{(n!)^2}{((n+1)!)^2}\right|\left|
\frac{\left|\gamma(2n+1,-\log( \mathrm{\mathsf{x}}))\right|}
{\left|\gamma(2n-1,-\log( \mathrm{\mathsf{x}}))\right|}\right|
=\lim_{n\rightarrow\infty} \left|\frac{1}{(n+1)^2}\right|
\log( \mathrm{\mathsf{x}})^2=0\;.
\end{equation}
However, since
$\lim_{ \mathrm{\mathsf{x}}\rightarrow\infty}\gamma(2n-1,-\log( \mathrm{\mathsf{x}}))=-\Gamma(2n-1)$,
and the sequence
\begin{equation}
\{a_n\}:=\frac{(-1)^n\Gamma(2n-1)}{\Gamma(n+1)^2}
\end{equation}
does not converge to zero, then $\overline{\pi_{2i}( \mathrm{\mathsf{x}})}$
diverges as $ \mathrm{\mathsf{x}}\rightarrow\infty$. Therefore, given the hypothesis of the
constrained gamma process for locating joint prime numbers, we
conclude that
\begin{equation}\label{infinite}
\lim_{ \mathrm{\mathsf{x}}\rightarrow\infty}\overline{\pi_{2i}( \mathrm{\mathsf{x}})}\rightarrow\infty
\;\;\;\;\forall \,i\in\N_+\;.
\end{equation}

This is reasonable for small $i$, but it seems questionable when $2i\rightarrow \mathrm{\mathsf{x}}-2$: Intuitively, it is hard to believe that there are an infinite number of prime doubles $(p,p+2i)$ if $i$ gets too large. On the other hand, technically $i$ must remain finite while $ \mathrm{\mathsf{x}}$ is allowed to go to infinity. So (\ref{infinite}) is possible if $C_{2i}$ remains relatively constant for all $i\in\N_+$, because then there are an infinite number of primes available for pairing. In other words, in the limit $ \mathrm{\mathsf{x}}\rightarrow\infty$, there are an infinite number of points  that prime doubles $(p,p+2i)$ (which are separated by a finite distance) can straddle.  Of course intuition is often flawed, and one should rigorously examine the assumption of simple normalization by $C_{2i}$.

Owing to its probabilistic foundation, the prime double hypothesis
cannot be confirmed unconditionally. However, given the success of the average prime counting
function $\overline{\pi_{1}( \mathrm{\mathsf{x}})}$, it appears plausible that the
hypothesis is correct; $C_{2i}$ notwithstanding. In particular, if we accept it for at least
$i=1$, then verification of the twin prime counting conjecture
follows immediately --- albeit \emph{conditionally} --- since the sum diverges with $ \mathrm{\mathsf{x}}$.

Of course, one might argue that the hypothesis is just an
alternative to the Hardy-Littlewood twin prime conjecture. However, the hypothesis is not asymptotic. With it we can
statistically verify the Goldbach conjecture:

\begin{proposition}
If the occurrence of prime doubles is a constrained gamma
process, then every even number greater than 2 is asymptotically
almost surely the sum of two primes.
\end{proposition}

\emph{Proof outline}: Assume the contrary. Then there exists a
$2 \mathrm{\mathsf{x}}$ that is not the sum of two primes. Clearly, $ \mathrm{\mathsf{x}}$
cannot be prime. Further, $ \mathrm{\mathsf{x}}$ cannot be `straddled' by a prime
double $(p,p+2i)$ with $p+i= \mathrm{\mathsf{x}}$  for some
$i\in\{1,\ldots, \mathrm{\mathsf{x}}-1\}$ since otherwise
$p+[p+2i]=( \mathrm{\mathsf{x}}-i)+[( \mathrm{\mathsf{x}}-i)+2i]=2 \mathrm{\mathsf{x}}$.

But according to the constrained gamma process hypothesis, the
probability density of prime doubles straddling the point $ \mathrm{\mathsf{x}}$
is given by the absolutely converging series
\begin{equation}
P_i({ \mathrm{\mathsf{x}}})=\frac{C_{2i}}{2i}\sum_{n=1}^\infty(-1)^{n}
\,\frac{\Delta_+(n,{ \mathrm{\mathsf{x}}})+\Delta_-(n,{ \mathrm{\mathsf{x}}})}{\Gamma(n+1)^2}
\end{equation}
where
\begin{equation}
\Delta_\pm(n, \mathrm{\mathsf{x}}):=\left[\pm\gamma(2n-1,-\log( \mathrm{\mathsf{x}}\pm1))
\mp\gamma(2n-1,-\log( \mathrm{\mathsf{x}}))\right]\;.
\end{equation}
So the expected number of prime doubles that straddle $ \mathrm{\mathsf{x}}$ is
given by
\begin{equation}
S({ \mathrm{\mathsf{x}}})
:=\sum_{i=1}^{ \mathrm{\mathsf{x}}-1}\overline{\pi_{2i}(2{ \mathrm{\mathsf{x}}})}\,P_i({ \mathrm{\mathsf{x}}})\;.
\end{equation}

Now, for sufficiently large $ \mathrm{\mathsf{x}}$ the expected number goes like
$S( \mathrm{\mathsf{x}})\sim  \mathrm{\mathsf{x}}/(\log^4( \mathrm{\mathsf{x}}))$. Moreover, since
$ \mathrm{\mathsf{x}}/(\log^4( \mathrm{\mathsf{x}}))$ is monotonically increasing for
sufficiently large $ \mathrm{\mathsf{x}}$, it only takes calculating $S( \mathrm{\mathsf{x}})$
for a few small $ \mathrm{\mathsf{x}}$ to see that $S( \mathrm{\mathsf{x}})>1$ for all
$ \mathrm{\mathsf{x}}$ sufficiently large. Since the probability that $ \mathrm{\mathsf{x}}$ is straddled by
at least one prime double is $1-e^{-S( \mathrm{\mathsf{x}})}$, we have a
contradiction asymptotically almost surely.$\QED$

One can check explicitly up to some sufficiently large
cut-off that the probability of a contradiction is
essentially almost sure. For example at $ \mathrm{\mathsf{x}}=10^9$ we find the
expected number of straddling prime doubles $S(\mathsf{x})>
29000$ where we used the \emph{under-estimate}
$\sum_iC_{2i}/i\approx 1$ for simplicity. So the probability that
the next even integer is not the sum of two primes is less than
about $10^{-12500}$. It is perhaps disconcerting that the conjecture
cannot be settled with certainty by this argument, but it is
comforting that the probability that it is false
--- beyond where one is willing to explicitly check --- decreases
exponentially like $e^{-\epsilon \, \mathrm{\mathsf{x}}/(\log( \mathrm{\mathsf{x}}))^4}$ with $\epsilon=O(1)$ a positive constant.

One final implication: Since the probability associated with prime
doubles only depends on the gap between them through $C_{2i}$, the
probability of twin primes at an interval $[ \mathrm{\mathsf{x}}-1, \mathrm{\mathsf{x}}+1]/2$ is
the joint distribution to use for the conditional probability of two
primes being separated by a gap. So the expected gap between prime
$p_1$ and $p_2$ given $p_1$ is $P(p_1+1)^{-1}$, and it is easy to
establish that $P(p)^{-1}\sim\log(p)^2$. Hence Cram\'{e}r's
conjecture is true \emph{on average} --- given the constrained gamma hypothesis.

\section{Conclusions}
Constrained dynamical systems were studied from a function space
perspective using newly developed functional integration tools. The
tools rely on the notions of conditional and conjugate integrators
--- the analogs of conditional and conjugate probability
distributions in Bayesian inference theory. These notions show the
well-known Gaussian functional integrals to be only part of the
picture: To describe constrained systems, one must be able to
manipulate functional integrals over constrained function spaces
using conjugate integrator families.

Applying the constrained functional integral concepts, well-known
results were re-derived efficiently at the functional level.
Additionally, the framework allowed construction of a model for
various counting functions associated with prime numbers that give
excellent numerical estimates and, hopefully, a basis for better
understanding prime distributions. The examples analyzed here point
to the utility of gamma and Poisson integrator families, but it is
likely that other probability distribution analogs will be useful.

No attempt was made to develop methods to calculate non-trivial
gamma functional integrals. That $\mathrm{Z}(\tau')$ is comprised of the
incomplete gamma function and it
is defined for complex parameters, points to considerable complexity.
Evidently the study of $\mathcal{D}\gamma$ is an involved but
important project. The perturbation expansion notwithstanding, the
gamma functional integral can be expected to yield new calculation
techniques.

It would be fruitful to extend the concepts developed in this
article beyond simple QM. In particular, the domain of $x$ and
$\tau$ can be altered in obvious ways to include quantum fields and
loops. Similarly, the domain of $X_a$ can be extended to include
matrix-valued functions
--- opening the door to matrix QM. Together with the complex
Gaussian integrator and the complex nature of the gamma integrator,
such extensions would appear to offer broad applicability and
significant potential.

\end{document}